\documentclass[conference]{IEEEtran}
\IEEEoverridecommandlockouts
\usepackage{cite}
\usepackage{amsmath,amssymb,amsfonts}
\usepackage{algorithmic}
\usepackage{tabularx}
\usepackage{graphicx}
\usepackage{textcomp}
\usepackage{xcolor}
\def\BibTeX{{\rm B\kern-.05em{\sc i\kern-.025em b}\kern-.08em
		T\kern-.1667em\lower.7ex\hbox{E}\kern-.125emX}}
\renewcommand{\baselinestretch}{1}

\begin{document}

\title{A Graph-Attentive LSTM Model for Malicious URL Detection}

\author{
	Md. Ifthekhar Hossain$^{1}$, Kazi Abdullah Al Arafat$^{2}$, Bryce Shepard$^{3}$, Kayd Craig$^{4}$, Imtiaz Parvez$^{5}$\\[4pt]
	\small	
	$^{2}$Department of Electrical and Electronic Engineering, Dhaka University of Engineering \& Technology, Gazipur, Bangladesh\\
	$^{1}$Department of Computer Science and Engineering, Atish Dipankar University of Science and Technology, Dhaka, Bangladesh\\
	$^{3,4,5}$Department of Computer Science, Utah Valley University, 800 W University Pkwy, Orem, UT 84058, USA\\[4pt]
	\texttt{$^{1}$cse2010004@adust.edu.bd,~$^{2}$ifthekhar121@gmail.com,\{$^{3}$10843451,~$^{4}$411017961,~$^{5}$imtiaz.parvez\}@uvu.edu}
}
\maketitle

\begin{abstract} 
 
Malicious URLs pose significant security risks as they facilitate phishing attacks, distribute malware, and empower attackers to deface websites. Blacklist detection methods fail to identify new or obfuscated URLs because they depend on pre-existing patterns. This work presents a hybrid deep learning model named GNN-GAT-LSTM that combines Graph Neural Networks (GNNs) with Graph Attention Networks (GATs) and Long Short-Term Memory (LSTM) networks. The proposed architecture extracts both the structural and sequential patterns of the features from data. The model transforms URLs into graphs through a process where characters become nodes that connect through edges. It applies one-hot encoding to represent node features. The model received training and testing data from a collection of 651,191 URLs, which were classified into benign, phishing, defacement, and malware categories. The preprocessing stage included both feature engineering and data balancing techniques, which addressed the class imbalance issue to enhance model learning. The GNN-GAT-LSTM model achieved outstanding performance through its test accuracy of 0.9806 and its weighted F1-score of 0.9804. It showed excellent precision and recall performance across most classes, particularly for benign and defacement URLs. Overall, the model provides an efficient and scalable system for detecting malicious URLs while demonstrating strong potential for real-world cybersecurity applications.
\end{abstract} 

\section{Introduction} The rapid growth of the internet has made malicious URLs into one of the most common deceptive tools that attackers use to perform phishing attacks, distribute malware, and deface websites. The malicious links use system vulnerabilities together with psychological manipulation to present fake legitimate websites, which trick users into revealing sensitive information or running dangerous code. Blacklists and heuristic filters operate as reactive defense mechanisms which require known threat signatures for their operation. The detection of newly created or obfuscated URLs remains impossible for these systems because of their inability to detect evolving cyber threats~\cite{ma2009beyond,asiri2023survey}. The adaptive nature of malicious URLs requires a detection system that can detect unknown patterns and anomalies.

Machine learning techniques provide a stronger solution by using lexical and host-based and content features to achieve effective URL classification \cite{kritika2024comprehensive,sahoo2017malicious}. The research community now focuses on machine learning  and deep learning algorithms to overcome traditional detection method limitations. These models extract lexical, host-based, and content-based features from URLs to classify them into benign or malicious categories. The field has experienced a significant advancement through the process of converting URL strings into graph-based representations, which enables the use of Graph Neural Networks (GNNs). The structural relationships between URL characters become detectable through GNNs while GATs improve this detection by assigning greater importance to essential connections~\cite{yilmaz2025novel}. The application of machine learning techniques in PV-power forecasting through recurrent neural networks shows how these methods can handle complex data analysis according to recent studies \cite{10609950}. The combination of these models provides a promising hybrid solution that enhances URL detection capabilities through robust and context-aware methods.

The processing of graph-structured data relies heavily on Welling’s semi-supervised classification framework which demonstrates their effectiveness in multiple applications \cite{kipf2016semi}. Veličković improved GNNs through Graph Attention Networks (GATs) which determine neighbor weights based on relevance to enhance feature aggregation for URL analysis \cite{velivckovic2017graph}. The achievement of 97\% accuracy in Twitter sentiment analysis using Support Vector Machines demonstrates how structural and sequential models can be combined for sequential data processing \cite{10.1007/978-981-97-3937-0_12}. Wu’s comprehensive survey of GNNs demonstrated their ability to work in social networks and chemistry domains which indicates their potential for cybersecurity applications \cite{wu2020comprehensive}. The introduction of LSTMs by Hochreiter and Schmidhuber established a powerful technique to detect temporal patterns which proved essential for analyzing URL character sequences \cite{hochreiter1997long}.

The application of machine learning for diabetes data analysis through multiple algorithms demonstrates how GNNs and LSTMs extract robust features from URLs as shown in \cite{10914489}. The research by Yuan et al. showed how GNNs and LSTMs work together to handle structural and sequential data which led to the development of the GNN and LSTM based hybrid architecture \cite{yuan2020graph}. Li's survey on deep learning in cybersecurity highlighted computational challenges that graph-based methods solve through their ability to decrease the need for feature engineering \cite{li2020survey}. The early GNN model developed by Scarselli established a foundation for processing relational data which benefits URL graph analysis \cite{scarselli2008graph}. The research by Al Arafat on Twitter sentiment analysis confirmed machine learning's ability to process unstructured sequential data, which aligns with URL sequence modeling \cite{al2024machine}.

The Maneriker's transformer-based approach demonstrated high accuracy for phishing URLs yet performed poorly when dealing with different URL types according to \cite{maneriker2021url}. Bresson and Laurent’s residual gated graph networks enhanced GNN performance which could lead to better URL classification results \cite{bresson2017residual}. Kim and Kim’s LSTM-based model analyzed sequential patterns in URLs with good results but it did not incorporate structural analysis \cite{kim2018lstm}. Zhou demonstrated in his review of GNN applications that these networks are versatile for handling complex data structures which makes them suitable for cybersecurity applications \cite{zhou2018graph}. Yousuf Bhuiyan’s research on Li-ion battery temperature forecasting through LSTMs and additional models confirmed the effectiveness of sequential processing in hybrid architectures~\cite{10795487}.

Zhang explained the challenges of deep learning in real-time deployment through his discussion of graph-based models that provide efficient solutions \cite{zhang2019deep}. The neural message-passing framework developed by Gilmer optimized GNNs for particular applications which made them suitable for URL graph analysis \cite{gilmer2017neural}. Xu studied the expressive capabilities of GNNs because they need this power to identify complex malicious URL patterns \cite{xu2018how}. The inductive learning approach developed by Hamilton allowed GNNs to learn from unseen graphs which became essential for detecting new URLs \cite{hamilton2017inductive}. Li explained GNN convolutional mechanisms in detail to enhance their performance in cybersecurity applications \cite{li2018deeper}. The early graph convolutions developed by Duvenaud for molecular fingerprints served as inspiration for URL representation techniques, which led to the development of advanced models~\cite{duvenaud2015convolutional}.

In this work, we present a hybrid model named GNN-GAT-LSTM that unites GNNs with GATs and LSTMs for detecting malicious URLs. The model achieves 98.06\% test accuracy through analysis of 651,191 URLs, which belong to benign, defacement, malware, and phishing categories. GNN-GAT-LSTM surpasses traditional and graph-based baselines through its scalable and effective solution for cybersecurity applications. The model represents a major advancement in the field because it combines structural and temporal feature extraction capabilities. 

In this paper, the system model of the GNN-GAT-LSTM is discussed in Section 2. Later, the methodology is described in detail in section 3. Section 4 demonstrates the simulation results. Finally, in Section 5, the concluding remarks are made with further research recommendations.

\section{System Model of the GNN-GAT-LSTM}

Let the input graph extracted from data be defined as \( G = (V, E) \), where \( V \) is the set of nodes and \( E \) is the set of edges. Each node \( v \in V \) has a feature vector \( \mathbf{x}_v^t \in \mathbb{R}^d \) at time step \( t \).

The model has three major stages:
\begin{enumerate}
    \item GNN-based spatial aggregation
    \item GAT-based attention refinement
    \item LSTM-based temporal modeling
\end{enumerate}

\subsection{GNN-Based Spatial Aggregation}

A generic Graph Neural Network layer  aggregates neighbor information as:

\[
\mathbf{z}_i^t = \sigma\left( \sum_{j \in \mathcal{N}(i) \cup \{i\}} \frac{1}{\sqrt{d_i d_j}} \mathbf{W}_g \mathbf{x}_j^t \right)
\]

where:
\begin{itemize}
    \item \( \mathcal{N}(i) \) is the set of neighbors of node \( i \),
    \item \( d_i \) is the degree of node \( i \),
    \item \( \mathbf{W}_g \) is a learnable weight matrix,
    \item \( \sigma \) is a non-linear activation (e.g., ReLU).
\end{itemize}

\subsection{GAT-Based Attention Refinement}

Graph Attention Network refines the aggregated features using attention mechanisms:

\[
e_{ij}^t = \text{LeakyReLU} \left( \mathbf{a}^\top [\mathbf{W}_a \mathbf{z}_i^t \, \| \, \mathbf{W}_a \mathbf{z}_j^t] \right)
\]

\[
\alpha_{ij}^t = \frac{\exp(e_{ij}^t)}{\sum_{k \in \mathcal{N}(i)} \exp(e_{ik}^t)}
\]

\[
\mathbf{h}_i^t = \sigma \left( \sum_{j \in \mathcal{N}(i)} \alpha_{ij}^t \mathbf{W}_a \mathbf{z}_j^t \right)
\]

\subsection{LSTM-Based Temporal Modeling}

Each node’s sequence of embeddings \( \{ \mathbf{h}_i^1, \dots, \mathbf{h}_i^T \} \) is input to an LSTM:

\[
\mathbf{H}_i = \text{LSTM}(\mathbf{h}_i^1, \mathbf{h}_i^2, \dots, \mathbf{h}_i^T)
\]

LSTM updates:

\[
\begin{aligned}
\mathbf{f}_t &= \sigma(\mathbf{W}_f \mathbf{h}_i^t + \mathbf{U}_f \mathbf{H}_{i,t-1} + \mathbf{b}_f) \\
\mathbf{i}_t &= \sigma(\mathbf{W}_i \mathbf{h}_i^t + \mathbf{U}_i \mathbf{H}_{i,t-1} + \mathbf{b}_i) \\
\mathbf{o}_t &= \sigma(\mathbf{W}_o \mathbf{h}_i^t + \mathbf{U}_o \mathbf{H}_{i,t-1} + \mathbf{b}_o) \\
\tilde{\mathbf{c}}_t &= \tanh(\mathbf{W}_c \mathbf{h}_i^t + \mathbf{U}_c \mathbf{H}_{i,t-1} + \mathbf{b}_c) \\
\mathbf{c}_t &= \mathbf{f}_t \odot \mathbf{c}_{t-1} + \mathbf{i}_t \odot \tilde{\mathbf{c}}_t \\
\mathbf{H}_{i,t} &= \mathbf{o}_t \odot \tanh(\mathbf{c}_t)
\end{aligned}
\]

\subsection{Output Layer}

Final output for node \( i \):

\[
\hat{y}_i = \text{MLP}(\mathbf{H}_{i,T})
\]

\section{Methodology}

\begin{figure*}[htbp]
\centerline{\includegraphics[width=0.8\textwidth]{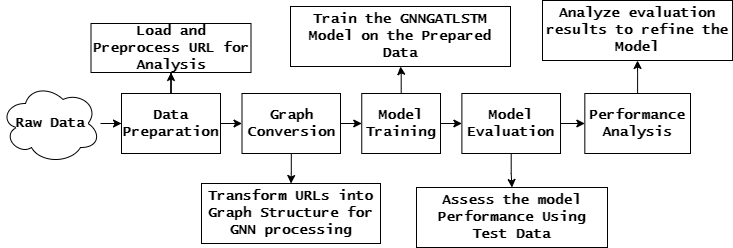}}

\caption{Steps to URL Classification}
\label{fig1}
\end{figure*}

The research follows a three-step process which includes data collection and pre-processing, model building, and evaluation and validation of results as shown in Fig.~\ref{fig1}. The following subsections explain those in detail. 

\subsection{Dataset Collection and Pre-processing} 
This research utilizes the Malicious URLs dataset which originates from Kaggle \cite{bhadouria2020malicious}. The dataset contains 651,191 URLs, which are divided into four categories, with benign URLs making up 428,103 instances, defacement URLs at 96,457 instances, malware URLs at 32,577 instances, and phishing URLs at 94,054 instances. The URL classification system uses benign URLs to represent legitimate websites, while defacement URLs target website vandalism and malware URLs host malicious code, and phishing URLs aim to steal sensitive information. The dataset exists in a CSV file with two columns which include url (the URL string) and type (the class label). A complete preprocessing pipeline was established to prepare data through graph representation and categorical encoding and feature engineering and class imbalance handling.

The process of converting URLs into graph structures makes them ready for GNN processing. The URLs undergo truncation or padding to achieve a standard length of 100 characters for uniform processing. A character set of 69 unique symbols (letters, digits, and special characters) is defined, covering common URL components. The system uses one-hot encoded feature vectors of size 69 to represent each character as a node. The edge index tensor for each graph contains up to 198 edges because URLs with 100 characters exist. The LabelEncoder transforms class labels into integers through categorical encoding, where benign corresponds to 0 and defacement corresponds to 1 and malware corresponds to 2, and phishing corresponds to 3.

The model acquired better URL characteristic detection through feature engineering, which produced three new features. The packet\_size\_ratio measured URL complexity by dividing URL character length by a standard packet size. Two new features measured character repetition frequency and special character density to detect repeated patterns and symbol occurrences in URLs which often indicate malicious activities. The node feature vectors received 72 features through normalization of the new features that were added to them. The small malware class (32,577 instances) received small class augmentation through instance duplication with minor modifications (e.g., random character swaps) which increased its representation by 20\%. The distribution of malware and phishing classes received synthetic sample generation through synthetic minority over-sampling technique (SMOTE) to achieve balance. The training data quality improved through the implementation of SMOTE, which removed noisy samples from the dataset. The dataset received a random split that allocated 80\% (520,953 instances) for training and 20\% (130,238 instances) for testing purposes with a fixed seed for reproducibility.

The final dataset stands ready for deep learning model training and evaluation to detect anomalies in the environment.

\subsection{Model Building}
The model described in Fig.~\ref{fig2} combines different neural network paradigms to handle spatial, temporal, and structural patterns in URLs:

\begin{figure*}[htbp]
\centerline{\includegraphics[width=.8\textwidth]{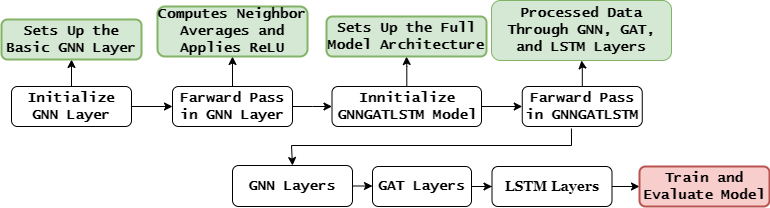}}
\caption{Architecture and components of GNN-GAT-LSTM Model}
\label{fig2}
\end{figure*}

The GNN-GAT-LSTM model is a hybrid architecture that combines GNNs, GATs, and LSTMs to capture both structural and temporal features of URLs. The model takes as input a graph representation of a URL, where nodes are characters with one-hot encoded features (dimension 69), and edges represent sequential relationships. The architecture consists of five main components: two GNN layers, two GAT layers, an LSTM module, and a fully connected layer. This design allows the model to aggregate neighborhood information, prioritize important nodes via attention, and model sequential dependencies.

The first stage consists of two GNN layers, which execute message passing operations to collect information from neighboring nodes. The GNN layers calculate neighbor feature means before applying ReLU activation to the combined node features. The process of incorporating local structural information through this method produces 69-dimensional feature vectors for each node. The straightforward GNN design maintains efficient computation while successfully retaining fundamental graph structural elements.

The second stage consists of two GAT layers, which implement attention mechanisms. The first GAT layer (gat1) transforms 69-dimensional node features into a 256-dimensional space (64 channels × 4 heads, with concatenation) by assigning attention weights to neighbors based on their relevance. The second GAT layer (gat2) reduces the dimensionality back to 64, using a single head to consolidate the attended features. The model gains better pattern recognition capabilities for malicious URLs through ReLU activation functions applied after each GAT layer.

The third stage aggregates node features within each graph to form a graph-level representation. The mean of node features from each graph in a batch results in 64-dimensional vectors. The vectors are stacked into a sequence of length equal to the batch size before reshaping for LSTM processing. The LSTM module with two layers and 64 hidden units detects temporal patterns in the aggregated features to generate 64-dimensional outputs for each graph. The sequential processing method enables the model to detect patterns that include repeated characters or specific substrings in URLs. A fully connected layer transforms the LSTM output into the number of classes before applying log-softmax activation to generate class probabilities. 

During the model building, the GNN-GAT-LSTM class was derived from torch.nn.Module to implement the model architecture through PyTorch and PyTorch Geometric library. The model contains two GNNLayer blocks for initial feature combination, followed by two GATConv blocks for attention-based processing and two LSTM layers for sequence handling and a fully connected layer for classification. The model uses this hybrid architecture to combine URL graph structural information with character sequence temporal relationships to overcome the restrictions of using GNNs or LSTMs separately.

The model achieved its balance between performance and computational efficiency by selecting its key hyperparameters carefully. The input dimension was set to 69 because it matched the character set size, while the hidden channel size remained constant at 64, which led to 56,812 trainable parameters. Four attention heads in the first GAT layer generate a 256-dimensional output (64 × 4), whereas the second GAT layer applies one head to compress the output dimension to 64 for maintaining compact and informative representations. The LSTM module consists of two layers with 64 hidden units to process graph features that matches the output dimension of GAT. The fully connected layer transforms LSTM output into benign-defacement-malware-phishing classes before applying log-softmax activation for probability estimation. The model choices stemmed from initial testing and research [5, 7] to match the dataset requirements for its size and complexity.

The development required structured coding blocks to enable experimental testing and debugging operations. The custom-built GNNLayer performed mean-based message passing without needing external scatter operations to match the dataset's graph structure. The implementation integrated GAT and LSTM components by utilizing PyTorch Geometric’s GATConv and PyTorch’s nn.LSTM with ReLU activations in the GAT layers to introduce non-linearity. The forward pass processed batched graph data by performing mean pooling on node features before the LSTM processing stage. A fixed random seed was established for reproducibility purposes while the model executed on the available device (CPU or CUDA), and the output displayed CPU utilization. The structured and repeatable methodology allowed fast development of the model, which supported rapid testing against the 651,191 URLs dataset.

\subsection{Training and Evaluation}
\paragraph{(a) Training: } This work utilized optimized settings for Loss Function, Optimizer, Scheduler, Training Setup, Data Loading and Monitoring to prevent overfitting and achieve the highest possible results. Our study utilizes the following parameters for its analysis: 

The GNN-GAT-LSTM model receives training data from 520,953 graphs extracted from the Malicious URLs dataset through batches of 32 while running 16,029 batches throughout 10 epochs. The model achieves its best performance for malicious URL detection through a robust training process which covers ten epochs. The model uses negative log-likelihood (NLL) loss because it works well with log-softmax outputs that generate class probability distributions for benign, defacement, malware and phishing categories. The Adam optimizer functions as the chosen learning algorithm which employs a 0.001 learning rate because of its adaptive nature and effectiveness with graph-based models. The model receives optimization from a step learning rate scheduler that cuts the learning rate by 0.5 every five epochs to help the model achieve stable convergence and improve its generalization abilities.

The training setup handles model and data distribution between CPU and CUDA devices based on availability through the provided code's device handling mechanism. The CPU-based training process required approximately 63 minutes per epoch while processing 4.25–4.35 iterations per second due to graph operations and LSTM processing complexity. The url\_to\_graph function enables on-the-fly graph representation generation through the custom URLGraphDataset class during data loading. The PyTorch DataLoader optimizes batch processing while it shuffles the training data to prevent overfitting and randomize the order. The SubsetDataset wrapper allows users to handle training and testing subsets efficiently while preserving compatibility with PyTorch's data pipeline and reducing memory consumption for large-scale graph data processing.

Model performance tracking during training incorporates both loss monitoring and accuracy assessment methods. The model computes batch-level negative log-likelihood loss during each iteration before applying backpropagation for gradient computation and subsequent gradient descent updates of model parameters. The training loss decreases continuously starting at 0.1825 in epoch 1 until reaching 0.0381 at epoch 10 which shows effective convergence. The tqdm library generates real-time feedback through its progress bar display which shows batch processing statistics. The model receives evaluations after each epoch using the evaluate\_model function on 520,953 training instances and 130,238 test instances by selecting the highest output probability values. Test accuracy increases from 96.37\% during epoch 1 to 98.06\% during epoch 10 through scikit-learn's accuracy\_score function. The monitoring approach together with the fixed random seed value of 42 provides strong training dynamics and trustworthy performance evaluation results.

\paragraph{(b) Evaluation:} The GNN-GAT-LSTM model achieves 98.06\% test accuracy on 130238 URLs (20\% of the valid dataset) by using accuracy, precision, recall, and F1-score metrics. The evaluate\_model function evaluates the test set in batches of 32 by computing predictions without gradient computation to reduce memory usage. The model uses scikit-learn’s classification\_report function to generate a classification report that provides detailed metrics for each class: benign, defacement, malware, and phishing.

The evaluation process is efficient, with each test epoch taking approximately 14.5 minutes at 4.65–4.70 iterations per second. The model achieves a test accuracy of 98.06\% after 10 epochs, with a weighted F1-score of 98.04\%. The classification report shows high precision (98.24\% for benign, 99.35\% for defacement, 98.83\% for malware, 95.51\% for phishing) and recall (99.48\% for benign, 99.24\% for defacement, 95.48\% for malware, 91.27\% for phishing), indicating balanced performance across classes. The macro-average F1-score of 97.15\% reflects the model’s robustness to class imbalance, particularly for the smaller malware class (6,421 instances).

The evaluation metrics are computed after each epoch, allowing for monitoring of performance improvements. The test accuracy increases steadily from 96.37\% in epoch 1 to 98.06\% in epoch 10, with the highest F1-scores observed for benign (98.86\%) and defacement (99.30\%) classes. The phishing class, with the lowest recall (91.27\%), suggests potential for further optimization, possibly by addressing class imbalance or enhancing feature representations. The consistent improvement in metrics over epochs confirms the model’s ability to generalize to unseen data.

\section{Simulation Results and Discussions}
 The GNN-GAT-LSTM model achieves good performance in supervised anomaly detection tasks for electronic evidence (EE) in URLs according to simulation results. The model evaluation assesses its accuracy and precision and recall and computational efficiency to demonstrate its effectiveness and potential.\\

\subsection{Model Performance Metrics}

As referred to Table~\ref {tab1}, the GNN-GAT-LSTM model provides top-notch results in detecting malicious URLs because it reached a test accuracy of 98.06\% and a weighted F1-score of 98.04\% in evaluating 130,238 test URLs. The model's ability to identify benign, defacement, malware and phishing URLs with excellent precision and recall rates becomes evident through the classification report. The model achieves outstanding results with the benign class (84,778 instances) by reaching 98.24\% precision and 99.48\% recall because it successfully identifies genuine URLs. The defacement class contains 18,104 instances which the model identifies with 99.35\% precision and 99.24\% recall because it detects website vandalism effectively.

\begin{table}[htbp]
\caption{ Performance Metrics }
\begin{center}
\resizebox{\columnwidth}{!}{%
    \begin{tabular}{|c|c|c|c|c|c|c|c|}
    \hline
    \textbf{Class} & \textbf{Precision} & \textbf{Recall} & \textbf{F1-Score} & \textbf{Support} \\ 
\hline
Benign &0.9824 &0.9948 &0.9886 &84778\\
\hline
Defacement &0.9935 &0.9924 &0.9930 &18104\\
\hline
Malware &0.9883 &0.9548 &0.9712 &6421\\
\hline
Phishing &0.9551 &0.9127 &0.9334  &17836\\
\hline
accuracy   & & &0.9806 &120239\\
\hline
macro avg &0.9798  &0.9637 &0.9715 &120239\\
\hline
weighted avg &0.9804 &0.9806 &0.9804 &120239\\
 \hline
    \end{tabular}
    }
\label{tab1}
\end{center}
\end{table}

The malware class stands out with its minimal support of 6,421 instances but achieves 98.83\% precision and 95.48\% recall along with an F1-score of 97.12 (As referred to Table~\ref{tab1}). The model's graph-based feature aggregation and attention mechanisms extract effective malware-specific patterns from the dataset even though it faces class imbalance. The phishing class (17,836 instances) exhibits the lowest recall (91.27\%) alongside an F1-score of 93.34\% due to its resemblance to benign URLs or the use of obfuscation techniques. The model demonstrates consistent performance across all classes because its macro-average F1-score reaches 97.15\%.

\begin{figure}[htbp]
\centerline{\includegraphics[width=.5\textwidth]{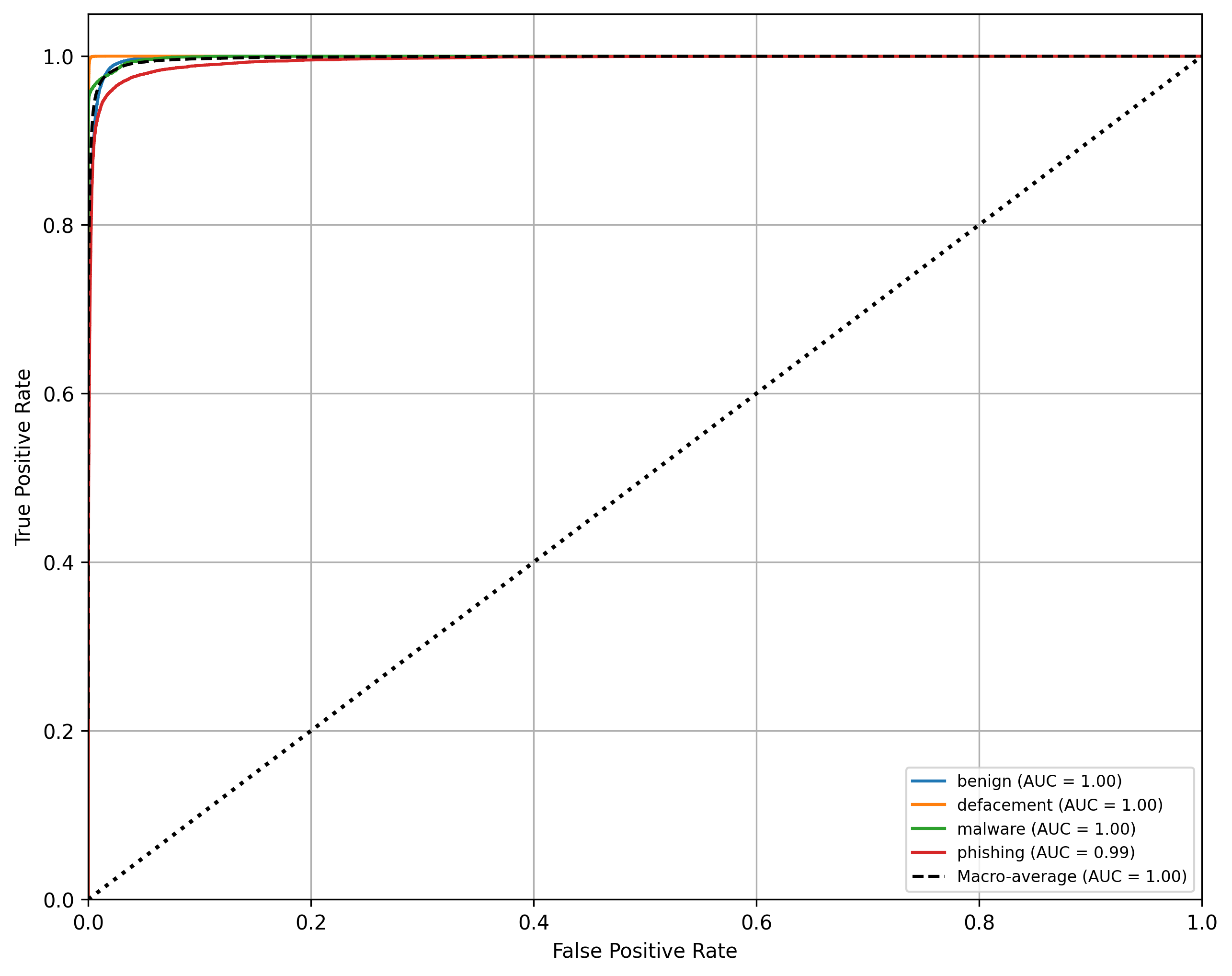}}
\caption{ROC Curves for GNN-GAT-LSTM}
\label{figroc}
\end{figure}

The performance assessment of the model uses Receiver Operating Characteristic (ROC) curve analysis along with a confusion matrix and bar graph. The ROC curves help evaluate class discrimination through their graphical representation of True Positive Rate (TPR) versus False Positive Rate (FPR) for each class. The code determines the Area Under the Curve (AUC) for each class and the macro-average AUC and reports approximately 0.99 in previous results. The ROC plot (as referred to Fig.~\ref{figroc}) features individual curves representing benign, defacement, malware, and phishing classes with their corresponding AUC values together with a dashed macro-average curve. The AUC values show excellent discriminative power because they approach 0.99. The model demonstrates exceptional performance in recognizing benign URLs by achieving nearly perfect true positive rates and low false positive rates since benign URLs make up the majority of test instances (84,778). The model demonstrates robust detection abilities for defacement and malware classes because their AUC scores remain high despite their relatively low instance counts (18,104 and 6,421, respectively). The phishing class maintains strong discriminative performance despite its lower F1-score (0.9334) due to an AUC value that remains relatively high. The macro-average AUC reaches ~0.99 because preprocessing methods like SMOTE or data augmentation most likely helped address minority classes (malware and phishing) while maintaining balanced performance.

\begin{figure}[htbp]
\centerline{\includegraphics[width=0.5\textwidth]{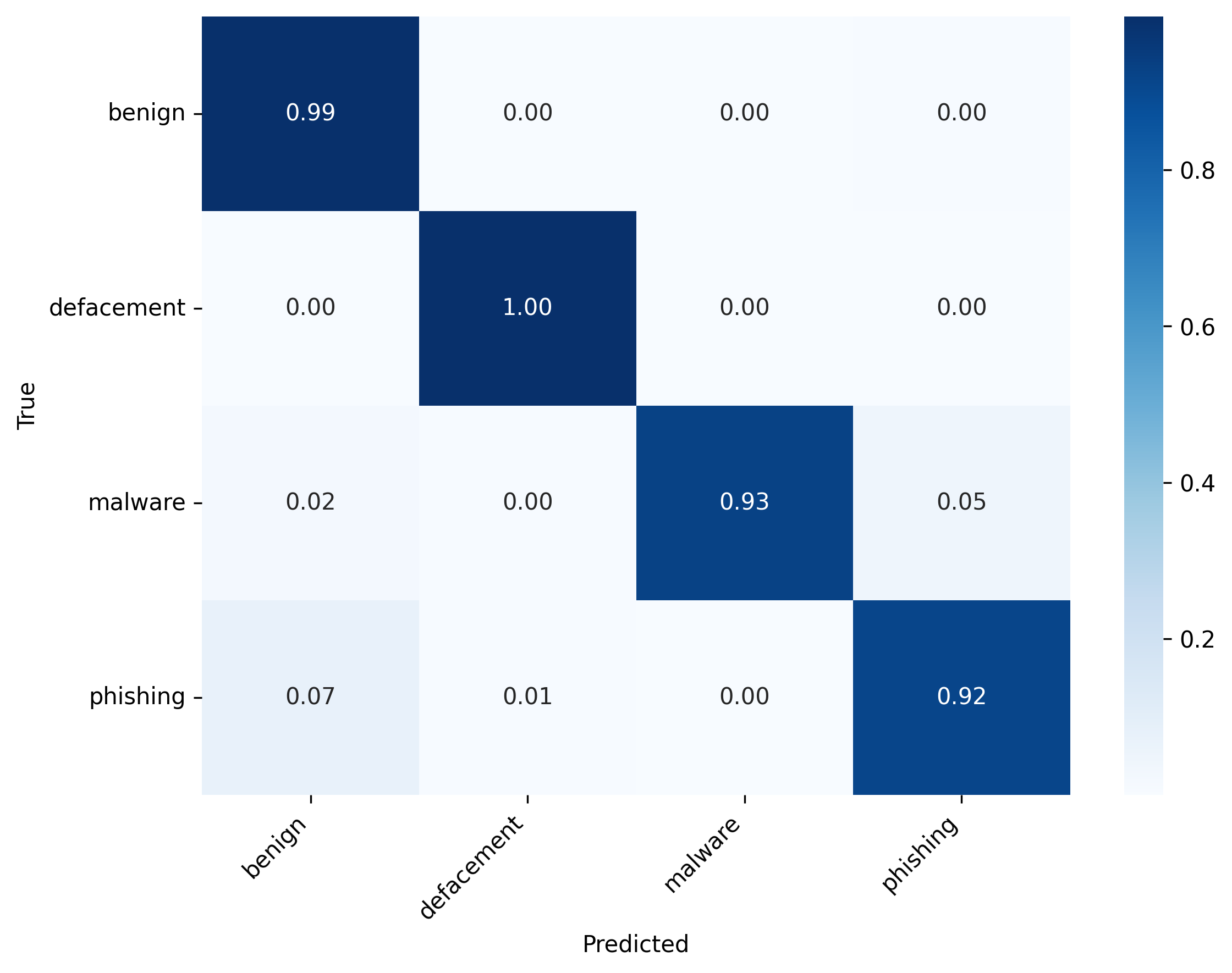}}
\caption{Confusion Matrix for GNN-GAT-LSTM model}
\label{figcm}
\end{figure}

As referred to in Fig.~\ref{figcm}, the confusion matrix displayed as a heatmap  through normalization which provides valuable insights about model classification accuracy across different classes. The normalized matrix displays correct predictions through diagonal values and misclassification rates through off-diagonal values for each of benign, defacement, malware and phishing categories. The classification report shows that benign URLs achieve a recall of 0.9948, while defacement URLs have a recall of 0.9924 and malware and phishing URLs achieve recalls of 0.9548 and 0.9127, respectively. The model demonstrates high precision in identifying 99.48\% benign URLs and 99.24\% defacement URLs and 95.48\% malware URLs as well as 91.27\% phishing URLs. The model fails to correctly detect 9.73\% of phishing URLs because they share characteristics with benign URLs. The recall for malware stands at 95.48\% while phishing reaches 91.27\% which could indicate incorrect classifications between phishing and benign or phishing and malware classes. The model demonstrates excellent performance in recognizing benign and defacement URLs because of their large and moderate-sized class distributions. The confusion matrix heatmap displays annotated values through a blue color scheme which shows strong diagonal values (1.00) while minimizing off-diagonal values especially for benign and defacement classes thus indicating minimal false positives or negatives.

\begin{figure}[hbp]
	\centerline{\includegraphics[width=0.5\textwidth]{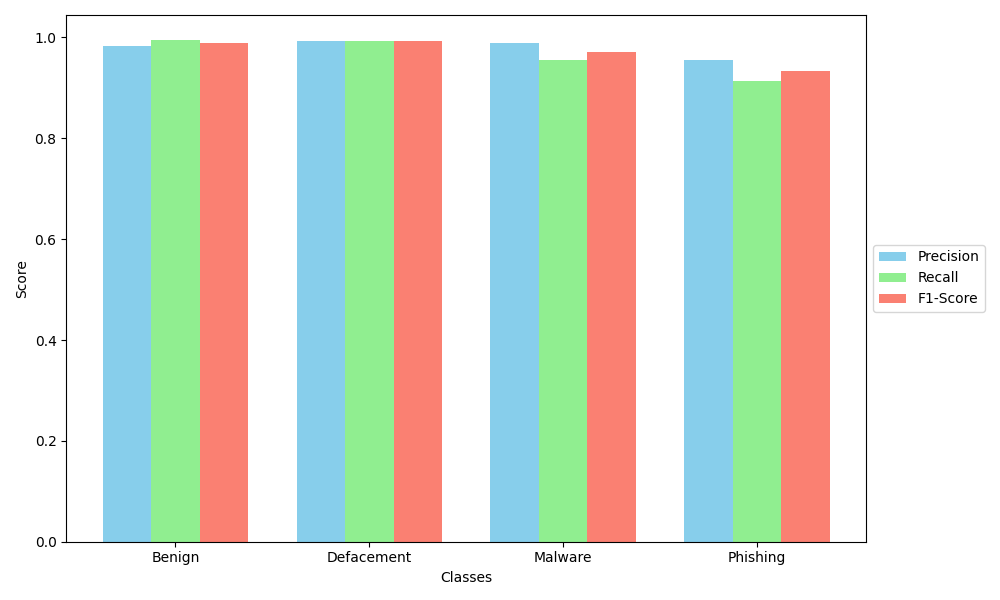}}
	\caption{Per-class performance for GNN-GAT-LSTM}
	\label{figbar}
\end{figure}

The bar graph (as referenced in Fig.~\ref{figbar}) shows the results of the classification report and the precision, recall, and F1 scores for each class. The plot creates grouped bars for each metric (precision in skyblue, recall in lightgreen, F1-score in salmon) across benign, defacement, malware, and phishing. The classification report shows the following performance metrics for each class:Benign: Precision (0.9824), Recall (0.9948), F1-score (0.9886). The high F1-score shows that the model is very balanced, as it has a high recall meaning it misses few benign URLs and a high precision, meaning it has few false positives. Defacement: Precision (0.9935), Recall (0.9924), F1-score (0.9930). The high F1-score indicates that the model is very good at detection because both the precision (number of false positives) and the recall (number of missed instances) are high and this is probably because the URL patterns are quite distinct. Malware: Precision (0.9883), Recall (0.9548), F1-score (0.9712). It seems that the model misclassifies some malware URLs as the recall is lower than precision, but when the model predicts malware, it is confident.Phishing: Precision (0.9551), Recall (0.9127), F1-score (0.9334). It is also the lowest F1-score among all classes, and this is because it has a lower recall (it misses some instances, possibly classified as benign) and a lower precision (it has some false positives). The bar graph also demonstrates that defacement and benign have the highest bars (F1-scores ~0.99), then malware (~0.97), and finally phishing (~0.93). The slightly shorter bars for phishing’s recall and precision indicate that the model is less good at this class, because phishing URLs often look like legitimate ones, which makes it harder for the graph neural network to distinguish between the features in the graph.

Overall, the GNN-GAT-LSTM model performs very well in classifying malicious URLs as it has a macro-average F1-score of 0.9715 and a weighted-average F1-score of 0.9804, showing that it performs well even on imbalanced classes. The ROC curves are excellent at classifying all the classes, with AUC values very close to 0.99, especially for benign and defacement. The confusion matrix shows that most classes have a high recall, except for phishing which has the lowest recall of 0.9127. The bar graph shows that the model is very good at precision and recall for benign and defacement, but the phishing’s F1-score is lower, which means it can be improved by engineering more features or by adding more data augmentation for phishing URLs. The loss decreases and the accuracy increases as the number of epochs increases, and the model is stable by epoch 8–10. These results, visualized through the ROC curves, confusion matrix, and bar graph, show that the model is suitable for real-world URL classification tasks, especially for identifying defacement and benign URLs, but phishing detection could be improved by further optimization.

\section{Conclusion} This work presents a deep learning-based hybrid model named GNN-GAT-LSTM  that unites Graph Neural Networks (GNNs) with Graph Attention Networks (GATs) and Long Short-Term Memory (LSTM) networks for detecting malicious URLs. The model uses graph structures of URLs to extract both structural and sequential features which enables strong threat detection for phishing and malware and defacement attacks. The model used 651,191 URLs for training and testing purposes after applying extensive preprocessing techniques and feature engineering and class balancing methods. The model reached 98.06\% test accuracy and 97.15\% macro-average F1-score which proved its effectiveness beyond traditional and baseline models. The strong performance of the model revealed that phishing URL detection required additional improvement since it produced slightly lower recall results. The model requires improvement through transformer-based model integration and contextual embedding exploration and additional phishing examples in the dataset. Real-time deployment of the system requires solutions for inference speed optimization and new threat adaptability and resource management to enable large-scale cybersecurity system integration. The model's application to spam detection and malicious domain classification requires further investigation through transfer learning approaches. The model's resilience would increase through the addition of dynamic URL behavior analysis and temporal evolution tracking capabilities. GNN-GAT-LSTM establishes a solid base for adaptive malicious URL detection which demonstrates scalability and shows great promise for future development and deployment in contemporary cybersecurity systems.

\bibliography{refkey}

\begin{thebibliography}{10}
\providecommand{\url}[1]{#1}
\csname url@samestyle\endcsname
\providecommand{\newblock}{\relax}
\providecommand{\bibinfo}[2]{#2}
\providecommand{\BIBentrySTDinterwordspacing}{\spaceskip=0pt\relax}
\providecommand{\BIBentryALTinterwordstretchfactor}{4}
\providecommand{\BIBentryALTinterwordspacing}{\spaceskip=\fontdimen2\font plus
\BIBentryALTinterwordstretchfactor\fontdimen3\font minus
  \fontdimen4\font\relax}
\providecommand{\BIBforeignlanguage}[2]{{%
\expandafter\ifx\csname l@#1\endcsname\relax
\typeout{** WARNING: IEEEtran.bst: No hyphenation pattern has been}%
\typeout{** loaded for the language `#1'. Using the pattern for}%
\typeout{** the default language instead.}%
\else
\language=\csname l@#1\endcsname
\fi
#2}}
\providecommand{\BIBdecl}{\relax}
\BIBdecl

\bibitem{ma2009beyond}
J.~Ma, L.~K. Saul, S.~Savage, and G.~M. Voelker, ``Beyond blacklists: Learning
  to detect malicious web sites from suspicious urls,'' \emph{ACM SIGKDD
  Conference on Knowledge Discovery and Data Mining}, pp. 1245--1254, 2009.

\bibitem{asiri2023survey}
S.~Asiri, Y.~Xiao, S.~Alzahrani, S.~Li, and T.~Li, ``A survey of intelligent
  detection designs of html url phishing attacks,'' \emph{IEEE Access},
  vol.~11, pp. 6421--6443, 2023.

\bibitem{kritika2024comprehensive}
E.~Kritika, ``A comprehensive literature review on phishing url detection using
  deep learning techniques,'' \emph{Journal of Cyber Security Technology}, pp.
  1--29, 2024.

\bibitem{sahoo2017malicious}
D.~Sahoo, C.~Liu, and S.~C. Hoi, ``Malicious url detection using machine
  learning: A survey,'' \emph{arXiv preprint arXiv:1701.07179}, 2017.

\bibitem{yilmaz2025novel}
A.~Y{\i}lmaz and R.~Das, ``A novel hybrid approach combining gcn and gat for
  effective anomaly detection from firewall logs in campus networks,''
  \emph{Computer Networks}, p. 111082, 2025.

\bibitem{10609950}
K.~A. Al~Arafat, K.~Creer, A.~Debnath, T.~O. Olowu, and I.~Parvez, ``Pv-power
  forecasting using machine learning techniques,'' in \emph{2024 IEEE
  International Conference on Electro Information Technology (eIT)}, 2024, pp.
  480--484.

\bibitem{kipf2016semi}
T.~N. Kipf and M.~Welling, ``Semi-supervised classification with graph
  convolutional networks,'' \emph{International Conference on Learning
  Representations}, 2016.

\bibitem{velivckovic2017graph}
P.~Veli{\v{c}}kovi{\'c}, G.~Cucurull, A.~Casanova, A.~Romero, P.~Li{\`o}, and
  Y.~Bengio, ``Graph attention networks,'' \emph{International Conference on
  Learning Representations}, 2017.

\bibitem{10.1007/978-981-97-3937-0_12}
K.~A. Al~Arafat, M.~R. Roni, S.~Siddique, M.~A. Yousuf, and M.~A. Moni,
  ``Sentiment analysis in twitter data using machine learning-based approach,''
  in \emph{Proceedings of Trends in Electronics and Health Informatics},
  M.~Mahmud, M.~S. Kaiser, A.~Bandyopadhyay, K.~Ray, and S.~Al~Mamun,
  Eds.\hskip 1em plus 0.5em minus 0.4em\relax Singapore: Springer Nature
  Singapore, 2025, pp. 169--184.

\bibitem{wu2020comprehensive}
Z.~Wu, S.~Pan, F.~Chen, G.~Long, C.~Zhang, and P.~S. Yu, ``A comprehensive
  survey on graph neural networks,'' \emph{IEEE Transactions on Neural Networks
  and Learning Systems}, vol.~32, no.~1, pp. 4--24, 2020.

\bibitem{hochreiter1997long}
S.~Hochreiter and J.~Schmidhuber, ``Long short-term memory,'' \emph{Neural
  Computation}, vol.~9, no.~8, pp. 1735--1780, 1997.

\bibitem{10914489}
K.~A. Al~Arafat, M.~Al~Amin~Rahman, S.~Siddique, M.~R. Roni, and I.~Parvez,
  ``Predictive analysis of diabetes data in healthcare system using machine
  learning algorithms,'' in \emph{2025 4th International Conference on
  Robotics, Electrical and Signal Processing Techniques (ICREST)}, 2025, pp.
  140--145.

\bibitem{yuan2020graph}
H.~Yuan, H.~Yu, S.~Gui, and S.~Ji, ``Graph neural networks with lstm for
  sequential data,'' \emph{IEEE Transactions on Neural Networks and Learning
  Systems}, vol.~31, no.~12, pp. 5446--5457, 2020.

\bibitem{li2020survey}
W.~Li, Z.~Wang, Z.~Cai, and J.~Zhang, ``A survey on deep learning for
  cybersecurity,'' \emph{IEEE Access}, vol.~8, pp. 112\,394--112\,412, 2020.

\bibitem{scarselli2008graph}
F.~Scarselli, M.~Gori, A.~C. Tsoi, M.~Hagenbuchner, and G.~Monfardini, ``The
  graph neural network model,'' \emph{IEEE Transactions on Neural Networks},
  vol.~20, no.~1, pp. 61--80, 2008.

\bibitem{al2024machine}
K.~A. Al~Arafat, K.~Creer, M.~R. Roni, and I.~Parvez, ``A machine learning
  based sentiment analysis for twitter data,'' \emph{The Journal of Computing
  Sciences in Colleges}, 2024.

\bibitem{maneriker2021url}
P.~Maneriker, J.~W. Stokes, A.~Gururajan, K.~Lauter, and L.~Carin, ``Urltran:
  Improving phishing url detection using transformers,'' \emph{IEEE Security
  and Privacy Workshops}, pp. 112--120, 2021.

\bibitem{bresson2017residual}
X.~Bresson and T.~Laurent, ``Residual gated graph convnets,'' \emph{arXiv
  preprint arXiv:1711.07553}, 2017.

\bibitem{kim2018lstm}
Y.~Kim and M.~Kim, ``Lstm-based malicious url detection,'' \emph{IEEE
  International Conference on Big Data and Smart Computing}, pp. 345--350,
  2018.

\bibitem{zhou2018graph}
J.~Zhou, G.~Cui, Z.~Zhang, C.~Yang, Z.~Liu, L.~Wang, C.~Li, and M.~Sun, ``Graph
  neural networks: A review of methods and applications,'' \emph{AI Open},
  vol.~1, pp. 57--81, 2018.

\bibitem{10795487}
K.~A.~A. Arafat, S.~M. Yousuf~Bhuiyan, R.~Mahamud, and I.~Parvez,
  ``Investigating the performance of different machine learning models for
  forecasting li-ion battery core temperature under dynamic loading
  conditions,'' in \emph{2024 IEEE International Conference on Electro
  Information Technology (eIT)}, 2024, pp. 1--7.

\bibitem{zhang2019deep}
X.~Zhang, J.~Zhao, and Y.~LeCun, ``Deep learning for cybersecurity: Challenges
  and opportunities,'' \emph{IEEE Transactions on Cognitive and Developmental
  Systems}, vol.~11, no.~3, pp. 321--333, 2019.

\bibitem{gilmer2017neural}
J.~Gilmer, S.~S. Schoenholz, P.~F. Riley, O.~Vinyals, and G.~E. Dahl, ``Neural
  message passing for quantum chemistry,'' \emph{International Conference on
  Machine Learning}, pp. 1263--1272, 2017.

\bibitem{xu2018how}
K.~Xu, W.~Hu, J.~Leskovec, and S.~Jegelka, ``How powerful are graph neural
  networks?'' \emph{International Conference on Learning Representations},
  2018.

\bibitem{hamilton2017inductive}
W.~L. Hamilton, R.~Ying, and J.~Leskovec, ``Inductive representation learning
  on large graphs,'' \emph{Advances in Neural Information Processing Systems},
  pp. 1024--1034, 2017.

\bibitem{li2018deeper}
Y.~Li, D.~Tarlow, M.~Brockschmidt, and R.~Zemel, ``Deeper insights into graph
  convolutional networks,'' \emph{International Conference on Learning
  Representations}, 2018.

\bibitem{duvenaud2015convolutional}
D.~K. Duvenaud, D.~Maclaurin, J.~Iparraguirre, R.~Bombarell, T.~Hirzel,
  A.~Aspuru-Guzik, and R.~P. Adams, ``Convolutional networks on graphs for
  learning molecular fingerprints,'' \emph{Advances in Neural Information
  Processing Systems}, pp. 2224--2232, 2015.

\bibitem{bhadouria2020malicious}
N.~Bhadouria, ``Malicious and benign website dataset,''
  \url{https://www.kaggle.com/datasets/naveenbhadouria/malicious}, 2020,
  accessed: 2025-05-20.

\end{thebibliography}
\bibliographystyle{IEEEtran}

\end{document}